\documentclass[11pt]{article}
\usepackage{moriond}

\usepackage{amsmath}
\usepackage{graphicx}

\def\be{\begin{equation}}
\def\ee{\end{equation}}
\def\bea{\begin{eqnarray}}
\def\eea{\end{eqnarray}}

\begin{document}
\vspace*{4cm}
\title{Explaining the LHC flavour anomalies}

\author{ANDREAS CRIVELLIN$^{*1}$, GIANCARLO D'AMBROSIO$^1$ and JULIAN HEECK$^2$}
\address{$^*$Speaker\\
$^1$CERN Theory Division, CH-1211 Geneva 23, Switzerland\\
$^2$Service de Physique Th\'eorique, Universit\'e Libre de Bruxelles, B-1050 Brussels, Belgium}

\maketitle\abstracts{The LHC observed deviations from the Standard Model (SM) in the flavour sector: LHCb found a $4.3\,\sigma$ discrepancy compared to the SM in $b\to s\mu\mu$ transitions and CMS reported a non-zero measurement of $h\to\mu\tau$ with a significance of $2.4\,\sigma$. Here we discuss how these deviations from the SM can be explained, focusing on two models with gauged $L_\mu-L_\tau$ symmetry. The first model contains two scalar doublets and vector-like quarks while the second one employs three scalar doublets but does not require vector-like fermions. In both models, interesting correlations between $b\to s\mu\mu$ transitions, $h\to\mu\tau$, and $\tau\to3\mu$ arise.}

\section{Introduction}
\label{intro}

The LHC completed the SM by discovering the Brout--Englert--Higgs particle~\cite{Aad:2012tfa,Chatrchyan:2012ufa}. While no significant direct evidence for physics beyond the SM has been found, the LHC did observe 'hints' for new physics (NP) in the flavor sector, which are sensitive to virtual effects of new particles and can be used as guidelines towards specific NP models: $h\to\mu\tau$, $B\to K^* \mu^+\mu^-$, $B_s\to \phi \mu^+\mu^-$ and $R(K)=B\to K \mu^+\mu^-/B\to K e^+e^-$. It is therefore interesting to examine if a specific NP model can explain these four anomalies simultaneously. In Refs.~\cite{Crivellin:2015mga,Crivellin:2015lwa}, two variants of such a model were presented, which we want to review here.

LHCb reported deviations from the SM predictions~\cite{Egede:2008uy,Altmannshofer:2008dz} (mainly in an angular observable called $P_5^\prime$~\cite{Descotes-Genon:2013vna}) in $B\to K^* \mu^+\mu^-$~\cite{Aaij:2013qta} with a significance of $2$--$3\,\sigma$. In addition, the measurement of $B_s\to\phi\mu\mu$ disagrees with the SM predictions by about $3\,\sigma$ \cite{Horgan:2013pva,Horgan:2015vla}. This discrepancy can be explained in a model independent approach by rather large contributions to the Wilson coefficient $C_9$~\cite{Descotes-Genon:2013wba,Altmannshofer:2013foa,Horgan:2013pva}, i.e.~an operator $(\overline{s}\gamma_\alpha P_L b)(\overline{\mu}\gamma^\alpha \mu)$, which can be achieved in models with an additional heavy neutral $Z^\prime$ gauge boson~\cite{Gauld:2013qba,Buras:2013dea,Altmannshofer:2014cfa,Niehoff:2015bfa,Sierra:2015fma}. Furthermore, LHCb~\cite{Aaij:2014ora} recently found indications for the violation of lepton flavour universality in $B$ meson decays
\begin{equation}
	R(K)=\frac{B\to K \mu^+\mu^-}{B\to K e^+e^-}=0.745^{+0.090}_{-0.074}\pm 0.036\,,
\end{equation}
which disagrees from the theoretically rather clean SM prediction $R_K^{\rm SM}=1.0003 \pm 0.0001$~\cite{Bobeth:2007dw} by $2.6\,\sigma$. A possible explanation is again a NP contributing to $C_9^{\mu\mu}$ involving muons, but not electrons~\cite{Alonso:2014csa,Hiller:2014yaa,Ghosh:2014awa}. Interestingly, the value for $C_9$ required to explain $R(K)$ is of the same order as the one required by $B\to K^* \mu^+\mu^-$ \cite{Hurth:2014vma,Altmannshofer:2014rta}. The global fit to the $b\to s\mu\mu$ data presented at this conference gives a $4.3\,\sigma$ better fit to data for the assumption of NP in $C_9^{\mu\mu}$ only, compared to the SM fit \cite{Altmannshofer:2015sma}.

Concerning Higgs decays, CMS measured a lepton-flavour violating (LFV) channel~\cite{Khachatryan:2015kon} 
\begin{equation}
	{\rm Br} [h\to\mu\tau] = \left( 0.84_{-0.37}^{+0.39} \right)\% \,,
	\label{h0taumuExp}
\end{equation}
which disagrees from the SM (where this decay is forbidden) by about $2.4 \,\sigma$. 
Most attempts to explain this decay rely on models with an extended Higgs sector~\cite{Dery:2014kxa,Campos:2014zaa,Celis:2014roa,Sierra:2014nqa,Lee:2014rba}. One particular interesting solution employs a two-Higgs-doublet model (2HDM) with gauged $L_\mu-L_\tau$~\cite{Heeck:2014qea}.

\section{The models}

Our models under consideration are multi-Higgs-doublet models with a gauged $U(1)_{L_\mu-L_\tau}$ symmetry~\cite{Heeck:2014qea}.\footnote{The abelian symmetry $U(1)_{L_\mu-L_\tau}$ is an anomaly-free global symmetry within the SM~\cite{He:1990pn}, and also a good zeroth-order approximation for neutrino mixing with a quasi-degenerate mass spectrum, predicting a maximal atmospheric and vanishing reactor neutrino mixing angle~\cite{Binetruy:1996cs}. Breaking $L_\mu-L_\tau$ is mandatory for a realistic neutrino sector, and such a breaking can also induce charged LFV processes, such as $\tau\to3\mu$~\cite{Dutta:1994dx,Heeck:2011wj} and $h\to \mu \tau$~\cite{Heeck:2014qea}.} The $L_\mu-L_\tau$ symmetry with the gauge coupling $g^\prime$ is broken spontaneously by the vacuum expectation value (VEV) of a scalar $\Phi$ with $Q^{\Phi}_{L_\mu-L_\tau}=1$, leading to the $Z'$ mass
\begin{equation}
m_{Z'} = \sqrt2 g' \langle\Phi\rangle \equiv g' v_\Phi\,,
\end{equation}
and Majorana masses for the right-handed neutrinos.\footnote{Neutrino masses arise via seesaw with close-to-maximal atmospheric mixing and quasi-degenerate masses~\cite{Heeck:2014qea}.}

In both models at least two Higgs doublets are introduced which break the electroweak symmetry: $\Psi_1$ with $Q^{\Psi_1}_{L_\mu-L_\tau}=-2$ and $\Psi_2$ with $Q^{\Psi_2}_{L_\mu-L_\tau}=0$. Therefore, $\Psi_2$ gives masses to quarks and leptons while $\Psi_1$ couples only off-diagonally to $\tau\mu$:
\begin{align}
\begin{split}
\mathcal{L}_Y \ \supset\ &-\overline{\ell}_f Y^\ell_{i}\delta_{fi} \Psi_2 e_i - \xi_{\tau\mu} \overline{\ell}_3 \Psi_1 e_2 -\overline{Q}_f Y^u_{fi} \tilde{\Psi}_2 u_i - \overline{Q}_f Y^d_{fi} \Psi_2 d_i + \mathrm{h.c.}
\end{split}
\label{eq:yukawas}
\end{align}
Here $Q$ ($\ell$) is the left-handed quark (lepton) doublet, $u$ ($e$) is the right-handed up quark (charged lepton) and $d$ the right-handed down quark while $i$ and $f$ label the three generations. The scalar potential is the one of a $U(1)$-invariant 2HDM~\cite{Branco:2011iw} with additional couplings to the SM-singlet $\Phi$, which most importantly generate the doublet-mixing term 
\begin{equation}
V(\Psi_1,\Psi_2,\Phi)\ \supset \ 2 \lambda \Phi^2 \Psi_2^\dagger \Psi_1 \to \lambda v_\Phi^2 \Psi_2^\dagger \Psi_1 \equiv m_3^2 \Psi_2^\dagger \Psi_1\,,\nonumber
\end{equation}
that induces a small vacuum expectation value for $\Psi_1$~\cite{Heeck:2014qea}. We define $\tan\beta = \langle \Psi_2\rangle/\langle \Psi_1\rangle$ and $\alpha$ is the usual mixing angle between the neutral CP-even components of $\Psi_1$ and $\Psi_2$ (see for example~\cite{Branco:2011iw}). We neglect the additional mixing of the CP-even scalars with Re$[\Phi]$.

\begin{figure*}
\includegraphics[width=0.43\textwidth]{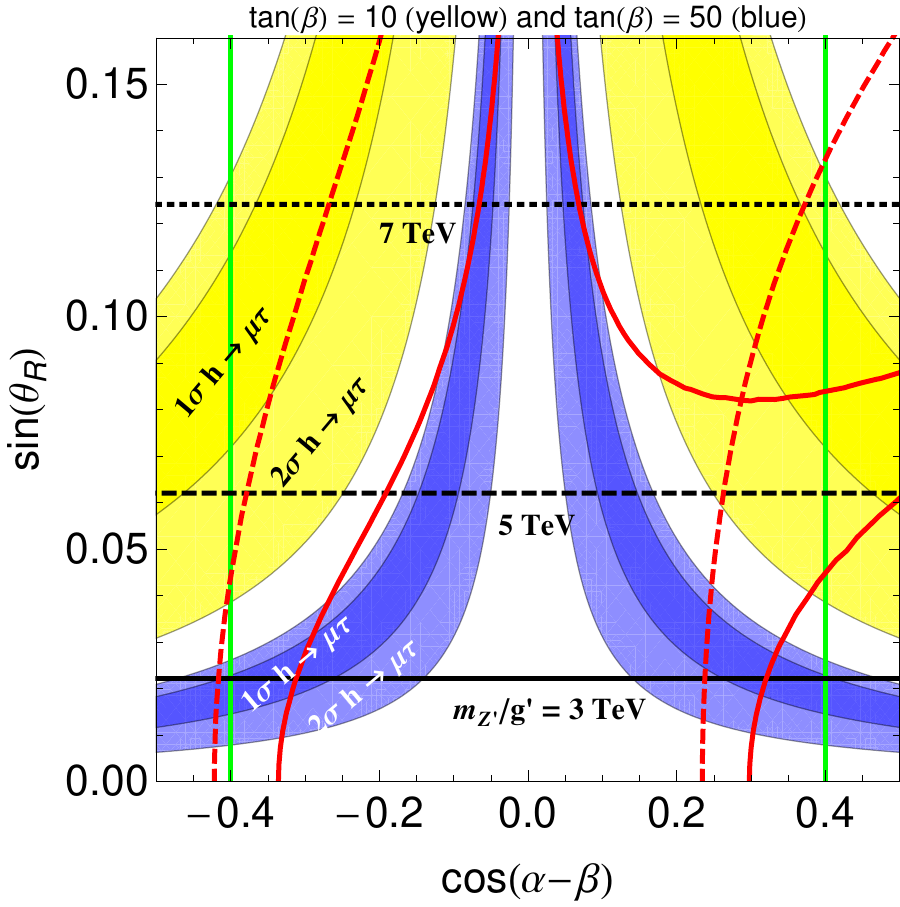} \hspace{3ex}
\includegraphics[width=0.42\textwidth]{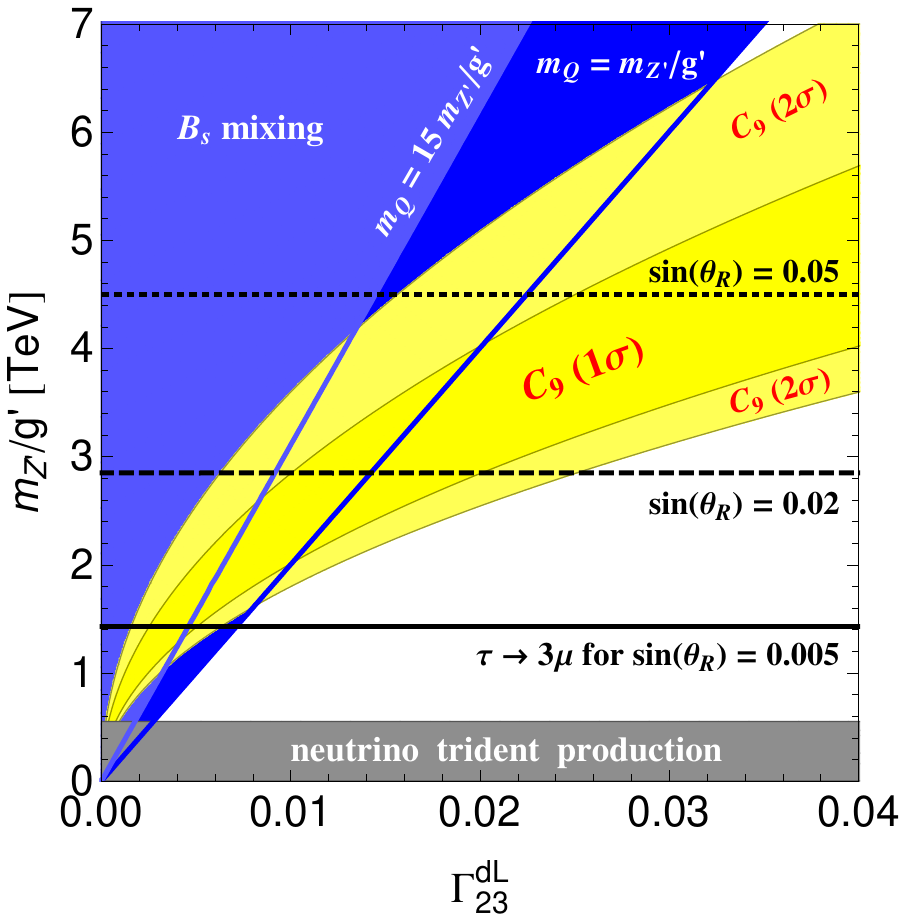}
\caption{Left: Allowed regions in the $\cos(\alpha-\beta)$--$\sin(\theta_R)$ plane. The blue (light blue) region corresponds to the $1\sigma$ ($2\sigma$) region of the CMS measurement of $h\to\mu\tau$ for $\tan \beta = 50$; yellow stands for $\tan \beta=10$. The (dashed) red contours mark deviations of $h\to\tau\tau$ by $10\%$ compared to the SM for $\tan\beta = 50$ ($10$). The vertical green lines illustrate the naive LHC limit $|\cos(\alpha-\beta)| \leq 0.4$, horizontal lines denote the $90$\%~C.L.~limit on $\tau\to 3\mu$ via $Z'$ exchange.
Right: Allowed regions for model~1 in the $\Gamma^{dL}_{23}$-$m_{Z'}/g'$ plane from $b\to s\mu^+\mu^-$ data (yellow) and $B_s$--$\overline{B}_s$ mixing (blue). For $B_s$--$\overline{B}_s$ mixing, (light) blue corresponds to ($m_Q=15 m_{Z'}/g'$) $m_Q=m_{Z'}/g'$. The horizontal lines denote the lower bounds on $m_{Z'}/g'$ from $\tau\to 3 \mu$ for $\sin(\theta_R)=0.05,\; 0.02,\; 0.005$. The gray region is excluded by neutrino trident production.}
\label{fig:HiggsPlot}
\end{figure*}

Quarks and gauge bosons have standard type-I 2HDM couplings to the scalars. The only deviations are in the lepton sector: while the Yukawa couplings $Y^\ell_{i}\delta_{fi}$ of $\Psi_2$ are forced to be diagonal due to the ${L_\mu-L_\tau}$ symmetry, $\xi_{\tau\mu}$ gives rise to an off-diagonal entry in the lepton mass matrix:
\begin{equation}
m^\ell_{fi}= \frac{v}{\sqrt{2}}\begin{pmatrix}
y_e\sin\beta &0&0\\
0& y_\mu \sin\beta& 0\\
0&\xi_{\tau\mu} \cos\beta& y_\tau \sin\beta
\end{pmatrix} .
\end{equation}
It is this $\tau$--$\mu$ entry that leads to the LFV couplings of $h$ and $Z'$ of interest to this letter. The lepton mass basis is obtained by simple rotations of $(\mu_R,\tau_R)$ and $(\mu_L,\tau_L)$ with the angles $\theta_R$ and $\theta_L$, respectively:
\begin{align}
 \sin \theta_R \simeq \frac{v}{\sqrt{2} m_\tau} \xi_{\tau\mu} \cos\beta \,, && \frac{\tan\theta_L}{\tan\theta_R} = \frac{m_\mu}{m_\tau} \ll 1\,.
\label{eq:thetaR}
\end{align} 
The angle $\theta_L$ is automatically small and will be neglected in the following.\footnote{Choosing $Q_{L_\mu-L_\tau}=+2$ for $\Psi_2$ would essentially exchange $\theta_L \leftrightarrow \theta_R$~\cite{Heeck:2014qea}, with little impact on our study.}
A non-vanishing angle $\theta_R$ not only gives rise to the LFV decay $h\to\mu\tau$ due to the coupling
\begin{equation}
\frac{m_\tau}{v}\frac{\cos(\alpha-\beta)}{\cos(\beta)\sin(\beta)}\sin(\theta_R)\cos(\theta_R) \bar\tau P_R\mu h\equiv \Gamma^{h}_{\tau\mu}\bar\tau P_R\mu h\,,\label{h0taumu}
\end{equation}
in the Lagrangian, but also leads to off-diagonal $Z'$ couplings to right-handed leptons
\begin{align}
g^\prime Z^\prime_\nu \, (\overline{\mu}, \overline{\tau})
\begin{pmatrix}
 \cos 2\theta_R& \sin 2\theta_R\\
\sin 2\theta_R& - \cos 2\theta_R
\end{pmatrix} \gamma^\nu P_R 
\begin{pmatrix}
 \mu\\
\tau
\end{pmatrix} ,
\end{align}
while the left-handed couplings are to a good approximation flavour conserving. $m_{Z'}/g'$ needs to be in the TeV range in order to suppress $\tau\to 3\mu$ if we want to explain $h\to\mu\tau$~\cite{Heeck:2014qea} (see Fig.~\ref{fig:HiggsPlot} (left)), which gives stronger bounds than neutrino trident production~\cite{Altmannshofer:2014cfa}.
In order to explain the observed anomalies in the $B$ meson decays, a coupling of the $Z'$ to quarks is required as well, not inherently part of $L_\mu-L_\tau$ models. 

\section{Model 1: vector-like quarks}

In order to couple the $Z^\prime$ to quarks we follow Ref.~\cite{Altmannshofer:2014cfa} and generate effective couplings via heavy vector-like quarks \cite{Langacker:2008yv} charged under $L_\mu-L_\tau$. As a result, the couplings of the $Z^\prime$ to quarks are in principle free parameters and can be parametrized as:
\begin{equation}
g'\left({{{\bar d}}_i}{\gamma ^\mu }{P_L}{{d}_j}{Z'_\mu }\Gamma_{ij}^{{d}L} + {{{\bar d}}}_i{\gamma ^\mu }{P_R}{{d}_j}{Z'_\mu }\Gamma_{ij}^{{d}R}\right)\,.
\end{equation}
In the limit of decoupled vector-like quarks with the quantum numbers of right-handed quarks, only $C_9$ is generated, giving a very good fit to data. The results are shown in the right plot of Fig.~\ref{fig:HiggsPlot} depicting that for small values of $\Gamma^L_{sb}$ and $\theta_R$, $b\to s\mu^+\mu^-$ data can be explained without violating bounds from $B_s$--$\overline{B}_s$ mixing or $\tau\to3\mu$. In the left plot of Fig.~\ref{fig:vevplot} the correlations of $b\to s\mu^+\mu^-$ and $h\to\mu\tau$ with $\tau\to3\mu$ are shown, depicting that consistency with $\tau\to3\mu$ requires large values of $\tan\beta$ (not being in conflict with any data as the decoupling limit is a type I model) and future searches for $\tau\to3\mu$ are promising to yield positive results. While this model predict tiny branching ratios for lepton-flavour-violating $B$ decays, these branching ratios can be sizable in generic $Z^\prime$ models in the presence of fine tuning in the $B_s$--$\overline{B}_s$ system~\cite{Crivellin:2015era}.

\section{Model 2: horizontal quark charges}

\begin{figure*}[t]
\includegraphics[width=0.41\textwidth]{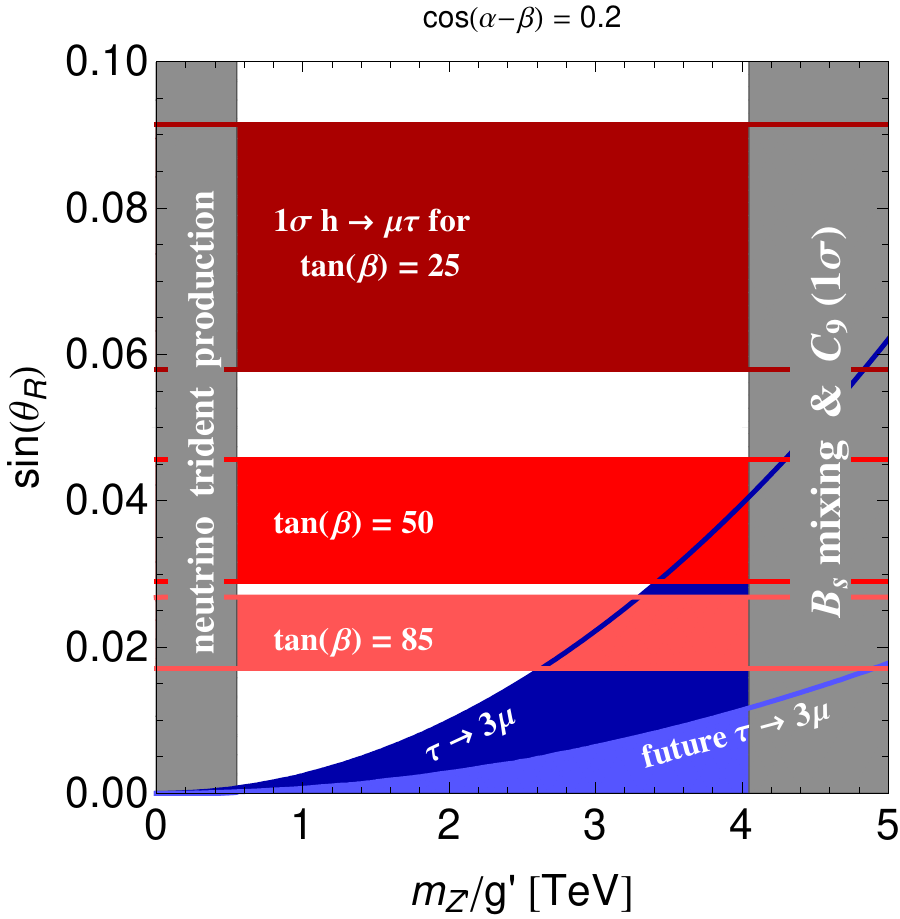}
\includegraphics[width=0.59\textwidth]{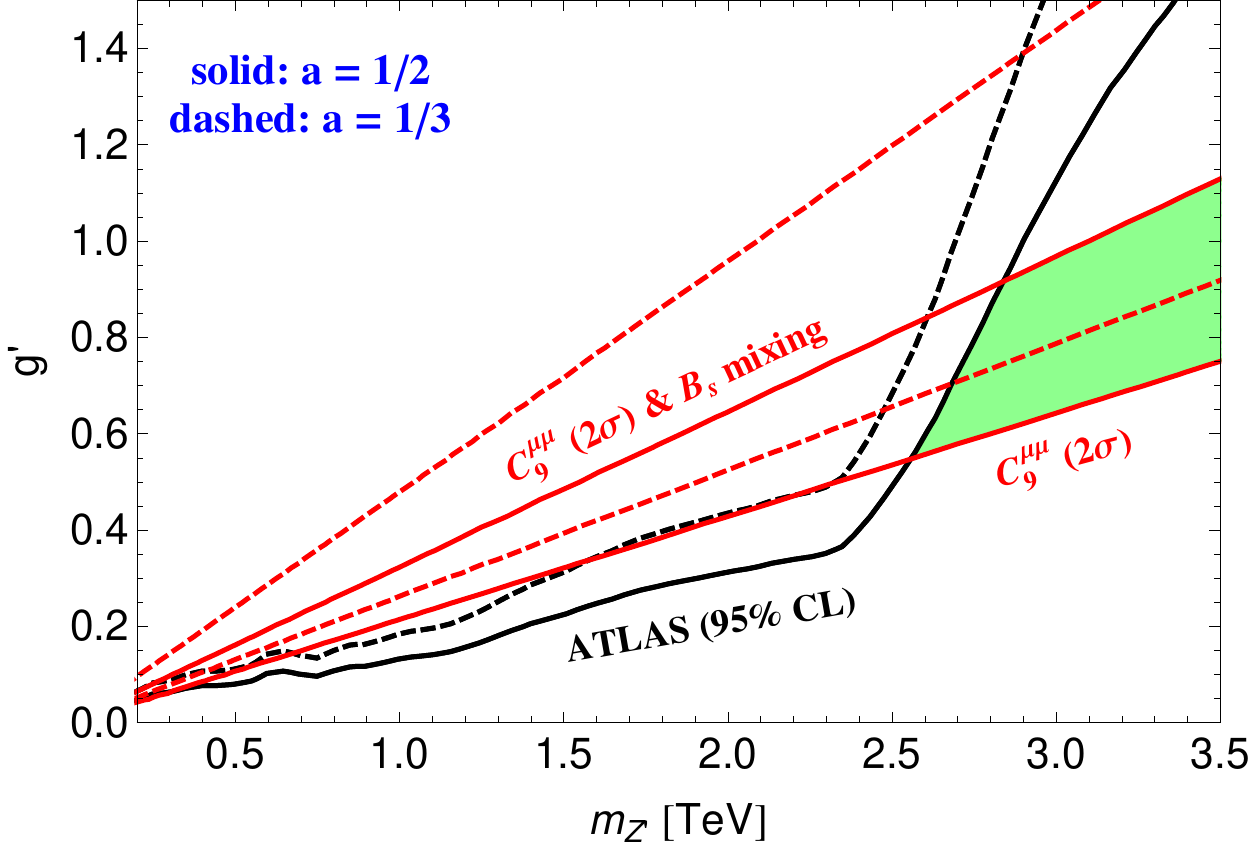}
\caption{Left: Allowed regions for model~1 in the $m_{Z'}/g'$--$\sin (\theta_R)$ plane: the horizontal stripes correspond to $h\to\mu\tau$ ($1\sigma$) for different $\tan\beta$ and $\cos(\alpha-\beta)=0.2$, (light) blue stands for (future) $\tau\to 3\mu$ limits at $90\%$~C.L. The gray regions are excluded by neutrino trident production or $B_s$--$\overline{B}_s$ mixing in combination with the $1\,\sigma$ range for $C_9$. 
Right: Limits for model~2 on $q\overline{q}\to Z' \to \mu\overline{\mu}$ from ATLAS (black, allowed region down right) and the $2\,\sigma$ limits on $C_9^{\mu\mu}$ to accommodate $b\to s\mu\mu$ data (allowed regions inside the red cone). Solid (dashed) lines are for $a=1/2$ ($1/3$). For $a =1/2$, the green shaded region is allowed (similar for $a= 1/3$ using the dashed bounds).\label{fig:vevplot}}
\end{figure*}

In order to avoid the introduction of vector-like quarks, one can assign flavour-dependent charges to baryons as well \cite{Crivellin:2015lwa}. Here, the first two generations should have the same charges in order to avoid very large effects in $K$--$\overline{K}$ or $D$--$\overline{D}$ mixing, generated otherwise unavoidably due to the breaking of the symmetry necessary to generate the measured Cabibbo angle of the CKM matrix. If we require in addition the absence of anomalies, we arrive at the following charge assignment for baryons	$Q'(B)= (-a,\,-a,\,2a )$, while leptons are still assigned $L_\mu-L_\tau$.
Here $a \in {\cal Q}$ is a free parameter of the model with important phenomenological implications. 
In this model, one additional Higgs doublet, which breaks the flavour symmetry in the quark sector, is required compared to the model with vector-like quarks. In case the mixing among the doublets is small, the correlations among $h\to\mu\tau$, $b\to s\mu^+\mu^-$ and $\tau\to 3\mu$ are similar as in the model with vector-like quarks discussed in the last subsection (left plot of Fig.~\ref{fig:vevplot}).

The low-energy phenomenology is rather similar to the one of the model with vector-like quarks (model~1), but the contributions to $B_s$--$\overline{B}_s$ mixing are directly correlated to $B_d$--$\overline{B}_d$ and $K$--$\overline{K}$ mixing, because all flavour violation is due to CKM factors. (These constraints are evaded for $a \leq 1$.) However, the implications concerning direct LHC searches are very different, as the $Z^\prime$ boson couples to quarks of the first generation and can be directly produced on-shell as a resonance in $p\bar p$ collisions. The resulting strong bounds are shown in right plot of Fig.~\ref{fig:vevplot}, where they are compared to the allowed regions from $B_s$--$\overline{B}_s$ mixing and $b\to s\mu^+\mu^-$ data for different values of $a$.

\section{Conclusions}
\label{conclusion}

In these proceedings we reviewed two variants of a model with a gauged $L_\mu-L_\tau$ symmetry which can explain all LHC anomalies in the flavour sector simultaneously: 1) a 2HDM with effective $Z^\prime\bar s b$ couplings induced by heavy vector-like quarks, 2) a 3HDM with horizontal charges for baryons. The models can account for the deviations from the SM in $b\to s\mu^+\mu^-$ data and $h\to\mu\tau$ simultaneously, giving also the desired effect in $R(K)$. Due to the small values of the $\tau$--$\mu$ mixing angle $\theta_R$, sufficient to account for $h\to\mu\tau$, the $Z'$ contributions to $\tau\to3\mu$ are not in conflict with present bounds for large $\tan\beta$ in wide rages of parameter space. Interestingly, $b\to s\mu^+\mu^-$ data combined with $B_s$--$\overline{B}_s$ put a upper limit on $m_{Z'}/g'$ resulting in a lower limit on $\tau\to3\mu$ if ${\rm Br}[h\to\mu\tau]\neq0$: for lower values of $\tan\beta$ the current experimental bounds are reached and future sensitivities will allow for a more detailed exploration of the allowed parameter space. The possible range for the $L_\mu-L_\tau$ breaking scale further implies the masses of the $Z'$ and the right-handed neutrinos to be at the TeV scale, potentially testable at the LHC with interesting additional consequences for LFV observables. While the low energy phenomenology of both models is rather similar, the variant with horizontal charges for baryons predicts sizable $Z^\prime$ production rates testable at the next LHC run. 

\section*{Acknowledgments}
We thank the organizers of the invitation to \emph{Moriond EW} and \emph{Moriond QCD} 2015 and for the opportunity to present these results. A.C. is supported by a Marie Curie Intra-European Fellowship of the European Community's 7th Framework Programme under contract number (PIEF-GA-2012-326948). 

\section*{References}


\begin{thebibliography}{99}

\bibitem{Aad:2012tfa}
  G.~Aad {\it et al.}  [ATLAS Collaboration],
  Phys.\ Lett.\ B {\bf 716} (2012) 1
  [arXiv:1207.7214].
	
\bibitem{Chatrchyan:2012ufa}
  S.~Chatrchyan {\it et al.}  [CMS Collaboration],
  Phys.\ Lett.\ B {\bf 716} (2012) 30
  [arXiv:1207.7235].
	
	
\bibitem{Crivellin:2015mga}
  A.~Crivellin, G.~D'Ambrosio and J.~Heeck,
  Phys.\ Rev.\ Lett.\  {\bf 114} (2015) 151801
  [arXiv:1501.00993].


\bibitem{Crivellin:2015lwa}
  A.~Crivellin, G.~D'Ambrosio and J.~Heeck,
  Phys.\ Rev.\ D {\bf 91} (2015)  075006
  [arXiv:1503.03477].


\bibitem{Egede:2008uy}
  U.~Egede, T.~Hurth, J.~Matias, M.~Ramon and W.~Reece,
  JHEP {\bf 0811} (2008) 032
  [arXiv:0807.2589].


\bibitem{Altmannshofer:2008dz}
  W.~Altmannshofer, P.~Ball, A.~Bharucha, A.~J.~Buras, D.~M.~Straub and M.~Wick,
  JHEP {\bf 0901} (2009) 019
  [arXiv:0811.1214].


\bibitem{Descotes-Genon:2013vna}
  S.~Descotes-Genon, T.~Hurth, J.~Matias and J.~Virto,
  JHEP {\bf 1305} (2013) 137
  [arXiv:1303.5794].


\bibitem{Aaij:2013qta}
  R.~Aaij {\it et al.}  [LHCb Collaboration],
  Phys.\ Rev.\ Lett.\  {\bf 111} (2013) 191801
  [arXiv:1308.1707].


\bibitem{Horgan:2013pva}
  R.~R.~Horgan, Z.~Liu, S.~Meinel and M.~Wingate,
  Phys.\ Rev.\ Lett.\  {\bf 112} (2014) 212003
  [arXiv:1310.3887].


\bibitem{Horgan:2015vla}
  R.~R.~Horgan, Z.~Liu, S.~Meinel and M.~Wingate,
  arXiv:1501.00367.


\bibitem{Descotes-Genon:2013wba}
  S.~Descotes-Genon, J.~Matias and J.~Virto,
  Phys.\ Rev.\ D {\bf 88} (2013) 074002
  [arXiv:1307.5683].


\bibitem{Altmannshofer:2013foa}
  W.~Altmannshofer and D.~M.~Straub,
  Eur.\ Phys.\ J.\ C {\bf 73} (2013) 2646
  [arXiv:1308.1501].


\bibitem{Gauld:2013qba}
  R.~Gauld, F.~Goertz and U.~Haisch,
  Phys.\ Rev.\ D {\bf 89} (2014) 015005
  [arXiv:1308.1959].


\bibitem{Buras:2013dea}
  A.~J.~Buras, F.~De Fazio and J.~Girrbach,
  JHEP {\bf 1402} (2014) 112
  [arXiv:1311.6729].


\bibitem{Altmannshofer:2014cfa}
  W.~Altmannshofer, S.~Gori, M.~Pospelov and I.~Yavin,
  Phys.\ Rev.\ D {\bf 89} (2014) 095033
  [arXiv:1403.1269].


\bibitem{Niehoff:2015bfa}
  C.~Niehoff, P.~Stangl and D.~M.~Straub,
  arXiv:1503.03865.


\bibitem{Sierra:2015fma}
  S.~D.~Aristizabal, F.~Staub and A.~Vicente,
  arXiv:1503.06077.


\bibitem{Aaij:2014ora}
  R.~Aaij {\it et al.}  [LHCb Collaboration],
  Phys.\ Rev.\ Lett.\  {\bf 113} (2014) 151601
  [arXiv:1406.6482].


\bibitem{Bobeth:2007dw}
  C.~Bobeth, G.~Hiller and G.~Piranishvili,
  JHEP {\bf 0712} (2007) 040
  [arXiv:0709.4174].


\bibitem{Alonso:2014csa}
  R.~Alonso, B.~Grinstein and J.~Martin Camalich,
  Phys.\ Rev.\ Lett.\  {\bf 113} (2014) 241802
  [arXiv:1407.7044].


\bibitem{Hiller:2014yaa}
  G.~Hiller and M.~Schmaltz,
  Phys.\ Rev.\ D {\bf 90} (2014) 054014
  [arXiv:1408.1627].


\bibitem{Ghosh:2014awa}
  D.~Ghosh, M.~Nardecchia and S.~A.~Renner,
  JHEP {\bf 1412} (2014) 131
  [arXiv:1408.4097].


\bibitem{Hurth:2014vma}
  T.~Hurth, F.~Mahmoudi and S.~Neshatpour,
  JHEP {\bf 1412} (2014) 053
  [arXiv:1410.4545].


\bibitem{Altmannshofer:2014rta}
  W.~Altmannshofer and D.~M.~Straub,
  arXiv:1411.3161.


\bibitem{Altmannshofer:2015sma}
  W.~Altmannshofer and D.~M.~Straub,
  arXiv:1503.06199.


\bibitem{Khachatryan:2015kon}
  V.~Khachatryan {\it et al.}  [CMS Collaboration],
  arXiv:1502.07400.


\bibitem{Dery:2014kxa}
  A.~Dery, A.~Efrati, Y.~Nir, Y.~Soreq and V.~Susič,
  Phys.\ Rev.\ D {\bf 90} (2014) 115022
  [arXiv:1408.1371].


\bibitem{Campos:2014zaa}
  M.~D.~Campos, A.~E.~C.~Hern{\'a}ndez, H.~P{\"a}s and E.~Schumacher,
  arXiv:1408.1652.


\bibitem{Celis:2014roa}
  A.~Celis, V.~Cirigliano and E.~Passemar,
  arXiv:1409.4439.


\bibitem{Sierra:2014nqa}
  D.~Aristizabal Sierra and A.~Vicente,
  Phys.\ Rev.\ D {\bf 90} (2014) 11,  115004
  [arXiv:1409.7690].


\bibitem{Lee:2014rba}
  C.~J.~Lee and J.~Tandean,
  JHEP {\bf 1504} (2015) 174
  [arXiv:1410.6803].


\bibitem{Heeck:2014qea}
  J.~Heeck, M.~Holthausen, W.~Rodejohann and Y.~Shimizu,
	Nucl.\ Phys.\ B {\bf 896} (2015) 281
  [arXiv:1412.3671].


\bibitem{He:1990pn}
  X.~G.~He, G.~C.~Joshi, H.~Lew and R.~R.~Volkas,
  Phys.\ Rev.\ D {\bf 43} (1991) 22.


\bibitem{Binetruy:1996cs}
  P.~Binetruy, S.~Lavignac, S.~T.~Petcov and P.~Ramond,
  Nucl.\ Phys.\ B {\bf 496} (1997) 3
  [hep-ph/9610481].


\bibitem{Dutta:1994dx}
  G.~Dutta, A.~S.~Joshipura and K.~B.~Vijaykumar,
  Phys.\ Rev.\ D {\bf 50} (1994) 2109
  [hep-ph/9405292].


\bibitem{Heeck:2011wj}
  J.~Heeck and W.~Rodejohann,
  Phys.\ Rev.\ D {\bf 84} (2011) 075007
  [arXiv:1107.5238].


\bibitem{Branco:2011iw}
  G.~C.~Branco, P.~M.~Ferreira, L.~Lavoura, M.~N.~Rebelo, M.~Sher and J.~P.~Silva,
  Phys.\ Rept.\  {\bf 516} (2012) 1
  [arXiv:1106.0034].


\bibitem{Langacker:2008yv}
  P.~Langacker,
  Rev.\ Mod.\ Phys.\  {\bf 81} (2009) 1199
  [arXiv:0801.1345].


\bibitem{Crivellin:2015era}
  A.~Crivellin, L.~Hofer, J.~Matias, U.~Nierste, S.~Pokorski and J.~Rosiek,
  arXiv:1504.07928.

\end{thebibliography}

\end{document}